\documentclass[aps,prl,reprint,groupedaddress,twocolumn]{revtex4-1}
\usepackage{amsmath}
\usepackage{siunitx}
\usepackage{graphicx}

\begin{document}


\title{}
\title{The Two Faces of Capacitance: New Interpretations for Electrical Impedance Measurements of Perovskite Solar Cells and Their Relation to Hysteresis}



\author{Daniel A. Jacobs}
\email[]{daniel.jacobs@anu.edu.au}
\author{Heping Shen}
\author{Florian Pfeffer}
\author{Jun Peng}
\author{Thomas P. White}
\author{Fiona J. Beck}
\author{Kylie R. Catchpole}
\affiliation{Research School of Engineering, The Australian National University, Canberra, ACT 2601, Australia.}


\date{\today}

\begin{abstract}
Perovskite solar cells are notorious for exhibiting transient behaviour not seen in conventional inorganic semiconductor devices. Significant inroads have been made into understanding this fact in terms of rapid ion migration, now a well-established property of the prototype photovoltaic perovskite MAPbI$_3$ and strongly implicated in the newer mixed compositions. Here we study the manifestations of ion migration in frequency-domain small-signal measurements, focusing on the popular technique of Electrical Impedance Spectroscopy (EIS). We provide new interpretations for a variety of previously puzzling features, including giant photo-induced low-frequency capacitance and negative capacitance in a variety of forms. We show that these apparently strange measurements can be rationalized by the splitting of AC current into two components, one associated with charge-storage, and the other with the quasi-steady-state recombination current of electrons and holes. The latter contribution to the capacitance can take either a positive or a negative sign, and is potentially very large when slow, voltage-sensitive processes such as ion migration are at play. Using numerical drift-diffusion semiconductor models, we show that giant photo-induced capacitance, inductive loop features, and low-frequency negative capacitance all emerge naturally as consequences of ion migration via its coupling to quasi-steady-state electron and hole currents. In doing so, we unify the understanding of EIS measurements with the comparably well-developed theory of rate dependent current-voltage (I-V) measurements in perovskite cells. Comparing the two techniques, we argue that EIS is more suitable for quantifying I-V hysteresis than conventional methods based on I-V sweeps, and demonstrate this application on a variety of cell types. 

\end{abstract}


\maketitle

\section{Introduction}

Halide perovskites have rapidly grown to prominence in the photovoltaic community due to their exceptional properties as thin-film photo-absorbers \cite{Stoumpos_Halide_2016,Docampo_A_2016,Correa-Baena_The_2017}. Starting with the prototype material of MAPbI$_3$, this class now encompasses an ever-growing compositional space based on substituting the halide anion, the organic cation and Pb in search of higher performance and stability. Record efficiencies of 22.7\% \cite{yang2017iodide} and climbing are already competitive with commercially established thin-film technologies, at least on a laboratory scale. Provided that stability issues can be adequately addressed, much excitement therefore surrounds the potential for these materials to enable a new generation of cheap and highly efficient solar cells, perhaps by partnering in a tandem architecture with an established technology such as the Si cell \cite{Boriskina_Roadmap_2016} or another relative newcomer such as CIGS \cite{Werner_Perovskite_2017}. 

In addition to their commercial prospects, halide perovskites have garnered intense scientific interest due to a number of unusual, and still partially unresolved, optoelectronic properties. This encompasses fundamental questions such as the nature of their bandgaps, questions of defect-tolerance, and the role of dynamic structural disorder, among other aspects \cite{egger2018remains}. On top of these already unusual features, the halide perovskites are unique in being both high-performing photovoltaic semiconductors and facile intrinsic-ion conductors at room temperature \cite{senocrate2017nature}. Materials exhibiting mixed ionic and electronic conductivity have been a topic of independent research for more than a half-century  \cite{wagner1975equations,gellings1997handbook,maier2004physical}, but the consequences of having a high ionic conductivity in photovoltaic devices are less well explored. Although the migration of intrinsic ionic defects is known in older photovoltaic materials such as CIGS \cite{gartsman1997direct}, it is not normal for ionic effects to manifest so clearly at ambient temperatures and on such short timescales as they do in I-V sweeps \cite{snaith2014anomalous} and other opto-electronic measurements of perovskite cells. In MAPbI$_3$ a large ionic conductivity is attributable to the mobility of I$^-$ in particular, likely through vacancy diffusion, although several lines of evidence point to the migration of MA$^+$ cations as well \cite{yuan2015photovoltaic}. On top of this intrinsic response, extrinsic ions from  selective contacts and metal electrodes could account for some of the observed transient phenomena in solar cell devices \cite{Li_Extrinsic_2017}. 

As perovskite technology matures it will become increasingly necessary to pin down the various aspects of device behaviour which remain unresolved in the literature at present. This problem is perhaps nowhere more pronounced than it is with frequency-domain measurements, particularly the popular technique of electrical impedance spectroscopy (EIS). As a characterization technique EIS has been called to give evidence on almost all major device properties, from built-in voltages and doping densities \cite{Laban_Depleted_2013,Liu_Electrical_2014}, to interface couplings \cite{Xu_Hole_2015,Chen_Hybrid_2015}, defect densities \cite{Samiee_Defect_2014,Duan_The_2015}, recombination mechanisms \cite{Shao_Origin_2014,Zaraza_Surface_2016,Contreras-Bernal_Origin_2017} and density of states \cite{Kim_Mechanism_2013}. Unfortunately, the relative ease of performing these measurements, as compared with the difficulty of verifying a given interpretation,  is such that there are now a great number of conflicting claims on this topic.  A related issue is that discrepant interpretations have arisen for frequency-domain measurements on the one hand, and for time-domain measurements including I-V hysteresis on the other, which should be related via the Fourier transform and not by a change in physical picture. Here we take steps towards resolving this issue by applying a class of ionic drift-diffusion models developed previously for studying I-V hysteresis to EIS measurements. These simulations provide considerable new insight into previously controversial observations regarding light-induced and negative capacitance. Such observations, which have been variously attributed to relatively exotic mechanisms such as polaron hopping \cite{Zarazua_Light_2016}, light-enhanced ion mobility \cite{kim2018large}, conduction through disordered capacitive networks \cite{Almond_An_2015}, and electrochemical reactions \cite{Zohar_Impedance_2016}, to name just a few, find a more unified and in some cases simple description in the ionic drift-diffusion (IDD) theory we address here.

We will begin in the first section with a summary of the IDD models used in this work, leaving a full specification, including detailed parameter choices, to the supplementary information. A proper understanding of light-induced and negative capacitance is greatly aided by recognizing separate contributions to the admittance, and therefore the measured capacitance, by charge storage on the one hand and quasi-steady state currents on the other: this distinction will be the described in the following section. In the ensuing sections we apply this theory to understand numerical simulations of EIS under the IDD model, focusing in particular on several phenomena of interest, namely, the giant photo-induced capacitance, so-called inductive loop features, and negative capacitance in low frequency measurements. We close with a summary of our interpretations and a specific application of these to the problem of quantifying I-V hysteresis.

\section*{Methods}
Our approach will treat perovskite cells as being foremost semiconductor devices, and accordingly take as a starting point the drift-diffusion equations of semiconductor physics. Mobile ions are added in the form of space charges subject to the drift-diffusion (Nernst-Planck) equation, and are therefore treated analogously to electrons and holes, except for being constrained to move only within the perovskite layer. For simplicity, no further mechanisms such as ionic generation-recombination, or interactions between ions and carriers beyond the electrostatic one, are assumed to be relevant in the first instance. Nonetheless, the problem is complex enough that a full numerical solution of the governing equations is necessary. The cost in generality of a numerical approach is compensated, in our view, by a relative freedom from constraints, including the ability to simulate multiple mobile ionic species, differing schemes of compensation, and arbitrary recombination models. The numerical approach also allows for a realistic treatment of the selective contact layers, including departures from flat-band conditions, which would not be readily accommodated by an analytic approach along the lines of refs. \cite{macdonald1973theory,macdonald1978theory,macdonald1992impedance}. Furthermore, the flexibility of this approach allows us to demonstrate that the IDD models are consistent with almost the full range of phenomena observed in EIS measurements, rather than being restricted to addressing just one or two cases of interest. 

\begin{figure}[h!]
	{\centering
		\includegraphics[width=3.5in]{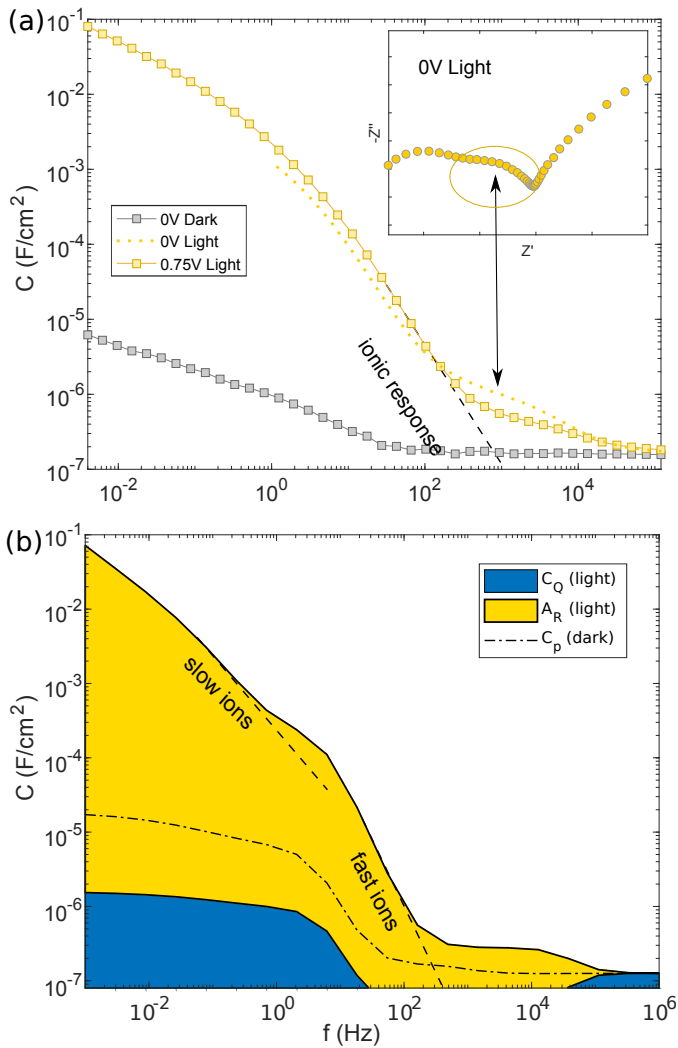}
	}
	\caption{(a) Experimental C-f measurements of a ``standard'' FTO/cp-TiO$_2$/mp-TiO$_2$/MAPbI$_3$/Spiro-OMeTAD/Au cell in darkness (grey) and under illumination (yellow). The dashed line indicates an extrapolation of the ionic contribution towards high frequencies. (a, inset) Nyquist plot at \SI{0}{\volt} under illumination, zoomed in on the high-frequency region (circled data corresponds to frequencies in the indicated portion of the C-f spectrum). (b) Cumulative area plot of a simulated C-f spectrum under illumination showing the relative contributions of phase-delayed recombination ($A_R$, yellow) and the more familiar charge-storage capacitance ($C_Q$, blue). Here two ionic species with diffusion constants \SI{1e-10}{\centi \metre \squared \per \second} and \SI{1e-13}{\centi \metre \squared \per  \second} were used to emulate the broad spectrum observed experimentally.  }
	\label{Fig1}
\end{figure}

Apart from their semiconductor parameters, IDD models are specified by the properties of the mobile ionic species plus those of the compensating charge. The most likely type of ionic disorder in MAPbI$_3$, on the basis of calculated enthalpies of formation, is partial MAI Schottky disorder with its very low energetic cost of \SI{0.08}{\electronvolt} per defect  \cite{Walsh_Self_2015}. If correct, this implies a high concentration of both MA$^+$ and I$^-$ vacancies, which sits well with evidence that both MA$^+$ and I$^-$ are mobile in MAPbI$_3$ \cite{yuan2015photovoltaic,yang2015significance}. Modelling the situation of pure MAI Schottky disorder is relatively straightforward, as it calls for two oppositely charged mobile species (the two vacancy types) in compensating amounts, but with unequal diffusion constants. All simulations in this paper utilize at least two ``ionic'' charge distributions, but for simplicity in most cases we set the slower species' diffusion constant to zero (the exception being Fig. \ref{Fig1}), as in previous studies \cite{van_Reenen_Modeling_2015,Richardson_Can_2016,Calado_Evidence_2016}. In this case the immobile species is assumed to have a uniform distribution throughout the absorber layer. Going further, one can also envisage situations where the compensation is imperfect, so that a net concentration of either positively or negatively charged defects results in an excess electron/hole density (net doping). Imperfect compensation of the mobile species leads to important effects, as we discuss in reference to Fig. \ref{Fig2} and the appearance of loop features. In that case, the imbalance is realized by adding a uniform density of acceptor dopants on top of the positive and negative ion distributions. All further details concerning our numerical models are included in the supplementary information.

\section*{Theory}
In an EIS measurement one is tasked with accounting for features in the linear response function, usually the impedance $Z(\omega)=\mathcal{F}\left[V(t) \right]/\mathcal{F}\left[I(t) \right]$ or equivalently admittance $Y=1/Z$, whose real and imaginary parts are related via the Kramers-Kronig relations \cite{EISMacdonald}. For this reason it is sufficient to give a physical account of only one component, as we shall do by focusing largely on the imaginary part. With reference to frequency-domain measurements the word ``capacitance'' is often used to denote what is strictly the AC or parallel capacitance $C=\operatorname{Im} \left[ Y \right]\omega^{-1}$. Given this convention, we will frequently refer to the AC capacitance as the measured capacitance in the following. Crucially, the AC capacitance differs from the \emph{static} notion of capacitance as the amount of charge associated with a change in (electrochemical) potential. From its definition, it is clear that the AC capacitance constitutes an inclusive measure of phase-delayed current oscillation, which may indeed stem from genuine charging currents, but can also be the result of delayed changes in the quasi-steady state current \cite{Jonscher_1986}. Such delayed changes in the current can in general occur for a wide variety of reasons.  For example, switching any diode-type device (a solar-cell or LED) into large forward bias will cause the device to heat up, over a timescale depending on its heat capacity and thermal coupling, causing a transient variation in the quasi-steady state current. In the frequency domain this would manifest as phase-delayed current and therefore show up in the measured capacitance, despite having nothing to do with stored charge. Acknowledging the inclusive nature of measured capacitance was essential to the proper interpretation of negative capacitance measurements in a wide range of conventional semiconductor devices \cite{Ershov_Negative_1998}. As we argue here, these distinctions play an outsize role in capacitance measurements of perovskite cells due to slow structural responses, in particular the rearrangement of mobile ions.

Considerations of current continuity make the difference between charge-storage capacitance and the general AC capacitance clear \cite{Laux_Revisiting_1999}, although experimentally these definitions are less accessible. A useful example to consider is the following continuity equation for electrons, with only generation and direct recombination included as source-sink terms for simplicity:
\[  e\frac{\partial n}{\partial t} = \frac{\mathrm{d }}{\mathrm{d }x}j_n+e \; g -e R \]
where $g$ and $R$ are the generation and recombination rates respectively, $n$ is the electron density, $j_n$ the electron current, and $e$ the electron charge. Considering just the first-order response to a sinusoidal voltage perturbation, this becomes (assuming constant generation)
\[ i \omega \, e \, \hat{n} = \frac{\mathrm{d }}{\mathrm{d }x}\hat{j}_n -e \hat{R} \]
in which a hat is used to denote the frequency-domain derivative (e.g. $n=n_0+\hat{n}\, e^{i \omega t} \mathrm{d}\, V$). In a device with carrier-selective contacts at both metallic terminals the above can be integrated to yield an expression for the admittance  \begin{align} Y &= \underbrace{ \, i\omega  \, \int  e \, \hat{n} \; \mathrm{dx}}_{Y_Q} +\underbrace{ \int  e\,    \hat{R} \; \mathrm{dx}}_{Y_R} \label{Ysplit}  \end{align}
in which each integral extends over the whole structure up to the metallic contacts (contributions from surface recombination and charges being implicitly included as delta sources). Interested readers are referred to ref. \cite{Laux_Revisiting_1999} for an analogous treatment in the context of p-n junctions, and to the supplementary information for further details regarding this decomposition. Equation (\ref{Ysplit}) shows that the measured admittance can be understood as comprising two terms relating to charge-storage ($Y_Q$) and recombination ($Y_R$) respectively. These components constitute the imaginary and real parts of $Y$ precisely when the carrier densities respond perfectly in-phase with the applied potential. In this case the conductivity $G=\operatorname{Re}\left[Y\right]$ relates strictly to the recombination current and the AC-capacitance $\operatorname{Im} \left[ Y \right]\omega^{-1}$ to stored charge.  For a variety of reasons carrier densities may fail to respond in phase however, meaning that $Y_R$ and $Y_Q$ are generically complex valued. This entails the possibility of non-zero contributions to the conductivity by $Y_{\text{Q}}$ and  perhaps more importantly  to the AC-capacitance by $Y_R$. Accordingly the AC capacitance under general circumstances is given by
\begin{eqnarray} C &=& C_Q + A_R \nonumber \\
&=&     \int   \operatorname{Re} \left[e \, \hat{ n} \right] \; \mathrm{dx} + \frac{1}{\omega} \int \operatorname{Im} \left[ e \, \hat{ R} \right] \; \text{dx} \; \label{A_R}  \end{eqnarray}
i.e. as the sum of a term relating to stored charge $C_Q$ and a contribution from  phase-delayed recombination $A_R$. The term $A_R$ has the dimensions of a capacitance, but this designation will be avoided to emphasize its distinct origin. Note that the stored charge is expressed here in terms of the electron density only; this is a consequence of our assumption that only electron current is allowed to flow past the electron-selective contact (both holes and ions being blocked). It does not preclude the contribution of ionic capacitance, which is included via the electron density needed to compensate accumulated ionic charge.

To obtain a better handle on $A_R$ we consider a device whose state responds to changes in the applied potential on a unique timescale $\tau$. For now this internal state may remain unspecified, but ultimately our concern will be with the state (distribution) of mobile ions in a perovksite absorber layer. The total recombination, for whatever reason, whether through its dependence on the carrier densities (hence internal field), or due to changes in the strength or number of recombination centres, is assumed to depend on this internal state. If the carriers are assumed to remain in quasi-steady state throughout the evolution (i.e. $\tau$ is much slower than the timescales of carrier dynamics), the total recombination current can be expressed as:

\begin{equation} R(t)=R_0 +\left[ (R_\infty -R_0)+ R_{\tau}  \exp \left( -t/\tau \right)\,\right] H (t) \Delta V \label{Rt} \end{equation}
where $R_{\tau}$ quantifies the transient or delayed response to a step-change in the  applied potential $V(t)=V_0+H(t) \Delta V$, with $H(t)$ the unit step function. $R_0 $ and $R_\infty$ respectively denote the recombination current immediately after and a long time after $t=0$.  Directly evaluating the Fourier transform of (\ref{Rt}), $\hat{R}=\mathcal{F}\left[\Delta R(t) \right]/\mathcal{F}\left[ H (t) \Delta V \right]$, and applying this in equation (\ref{A_R}) yields a Debye-type expression for $A_R$: 

\begin{equation}  A_R = \frac{\tau  \;  R_{\tau}}{1+(\omega \tau)^2} \;. \label{Cdebye} \end{equation}

A positive contribution to the AC capacitance will therefore result when recombination is initially large but decreased by the process causing delay ($ R_{\tau}>0$), and negatively to the capacitance when recombination increases over time ($ R_{\tau}<0$). This provides a particularly natural way for negative capacitance to arise in circumstances where the recombination current is significant compared to the charging current measured by $C_Q$ (see ref. \cite{Ershov_Negative_1998} for an extensive list of examples). Apart from its ready flexibility in sign, $A_R$ is also distinguished from the charge-storage capacitance by its explicit frequency dependence. In equation (\ref{Cdebye}) this results in a low-frequency limit that is linearly proportional to the timescale of relaxation $\tau$. Since there is in principle no limit to these timescales when slow agents such as migrating ions are present, the $A_R$ contribution can be arbitrarily large. Indeed, it is notable that replacing the Debye relaxation in equation (\ref{Cdebye}) with a Cole-Cole response, commonly used to model a broadened distribution of timescales parametrized by a factor $0<\alpha<1$ ($\alpha=1$ being the un-broadened case) \cite{EISMacdonald}, leads to $A_R$ capacitances which diverge as $\omega^{\alpha-1}$ at low frequencies.

In perovskite cells, mobile ions are present at a sufficient density to have a screening effect on the internal field \cite{Tress_Understanding_2015}, and respond slowly to changes in the applied potential. Through this coupling ions affect the collection probability of photogenerated charges, or in other words the recombination current, since collection is determined by total generation (a constant) minus recombination. Even without illumination ions can modulate the carrier density profiles to change the net rate of recombination. The slow response of mobile ions to variations in the applied potential therefore leads to transient effects in the quasi-steady-state recombination current, which is the meaning of ``phase-delayed recombination'' in subsequent discussion. It is important to note that even though the transient process of ion redistribution is rate-determined by a charging processes, namely accumulation at the device's internal interfaces, the capacitance measured in EIS is liable to be dominated by the contribution of phase-delayed recombination $A_R$ over that of the charging current $C_Q$ whenever carrier densities are sufficiently high, despite the latter contribution being always present.

In the following we apply numerical simulations to address the interpretation of EIS/capacitance measurements applied to perovskite cells, particularly those with the ``standard structure'' based on the use of TiO$_2$ as the electron selective contact (ETL) and Spiro-OMeTAD as the hole selective layer (HTL), since these have been subject to the most intense study to date. EIS measurements of such cells typically exhibit a high frequency response, associated with parallel-plate type charging, and a low-frequency response due to interfacial ion accumulation. Occasionally, distinct features appear in small-signal measurements at intermediate frequencies (1-100 Hz) as well \cite{Guerrero_Properties_2016,Pockett_Microseconds_2017}. We will consider the role of phase-delayed recombination, as defined by equation (\ref{A_R}), with reference to each of these features, starting with the low and high frequency capacitance under illumination, before moving to the more exotic intermediate features and other manifestations of negative capacitance. High-frequency capacitance-voltage (C-V) measurements under dark conditions are studied as a further check on the IDD theory. We close with a discussion about the relationship between EIS measurements and I-V hysteresis, and their potential application in quantifying the latter. 

\section*{Photo-induced Capacitance}

First we consider EIS measurements under illumination. An experimental C-f spectra of a standard MAPbI$_3$ cell is shown in Fig. \ref{Fig1}a, taken with and without illumination provided by a \SI{630}{\nano \metre} LED array at approximately 1 sun intensity. In the low-frequency part a single feature is apparent below \SI{1}{\kilo \hertz} under illumination and at somewhat lower frequencies in the dark (no shift in the characteristic timescale is evident in the Bode plot however). The low-frequency feature has a characteristic frequency of $\approx$ \SI{0.1}{\hertz} under illumination, as determined from the Bode plot, which is similar to timescales of I-V hysteresis for this cell type. A now considerable body of evidence points very strongly towards ion migration being the root cause of I-V hysteresis (see refs. \cite{Yuan_Ion_2016,saliba2017perovskite,tress2017metal} for reviews), with I$^-$ vacancies being the most likely culprit in MAPbI$_3$ \cite{Li_Iodine_2016,Meloni_Ionic_2016}. Given that a low-frequency EIS measurement is, in essence at least, an I-V measurement with a smaller amplitude and smoother waveform, the broad low-frequency feature can be attributed with reasonable confidence to ion migration. The dramatic enhancement of low-frequency capacitance observed when the cell is measured under illumination is therefore a likely manifestation of the familiar I-V hysteresis, simply on the basis of timescales. Since hysteresis is understood via the effect of ion populations on carrier collection, or recombination rates, it is natural to posit that the light-induced capacitance is explained by a large $A_R$ contribution according to equation (\ref{A_R}). As a check on this theory we have performed IDD simulations of this experiment, shown in Fig. \ref{Fig1}b. To roughly emulate the extreme broadening of the experimental low-frequency feature (which rises over 4 decades)  we have included two ionic species (fast and slow) in the simulation, although it is not clear how many distinct species are actually relevant in the device physics of MAPbI$_3$. Broadening of the experimental spectrum could alternatively result from lateral variations in the RC constant of a single ionic species, or from diffusion along irregular pathways (e.g. grain boundaries) with variable activation energies. Regardless, inclusion of a slower species in the simulation results in enormous values of the low-frequency capacitance of order \SI{1e-1}{ \farad \per \centi \metre \squared} due almost solely to the contribution from phase-delayed recombination ($A_R$), values which are similar in magnitude to those seen in our measurement. The massive contribution from $A_R$ is found to occur generically (that is for almost all parameter choices that include a slow-moving ionic species) with slower ions leading to a larger low-frequency limit in accordance with the scaling in equation (\ref{Cdebye}).

\begin{figure}[h!]
	{\centering
		\includegraphics[width=3.5in]{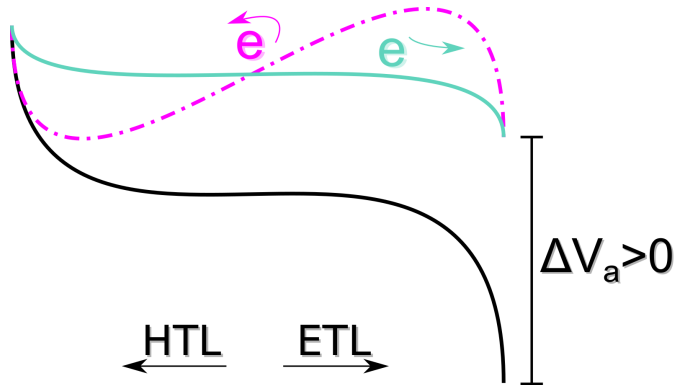}
	}
	\caption{Schematic of the evolution in bulk potential within the absorber layer of a perovskite cell following the application of forward bias. The potential is shown in steady state before the step to forward bias (black), immediately after the potential is applied (purple) and after ionic redistribution has occurred (cyan). }
	\label{Arschem}
\end{figure}

According to equation (\ref{A_R}) a positive $A_R$ contribution to the capacitance follows from $R_{\tau}>0$. This signifies a recombination current that decreases as a function of time following a step to forward bias (equation (\ref{Rt})), which may appear counter-intuitive. This can be understood physically by recognizing that the forward bias potential is initially exerted over the entire perovskite bulk (assuming low electron and hole densities), whereas after ionic screening has taken place the applied potential drops only near the contact interfaces (Fig. \ref{Arschem}). Experimentally $R_{\tau}>0$ is the trend observed in step-transient measurements \cite{Unger_Hysteresis_2014}, wherein moving to forward bias is found to result in temporarily reduced photocurrent that rises over time (suggesting a recombination rate that decreases over time, i.e. $R_{\tau}>0$ ). In EIS measurements, $R_{\tau}>0$ corresponds to a larger impedance at low frequencies which is the generally observed trend. Conceivably the opposite case of $R_{\tau}<0$ could also occur as a result of ionic redistribution, which would result in negative $A_R$ contributions. Such cases are considered in the sections below where we address loop features and other observations of negative capacitance. 

\section{Theories of Capacitance in Perovskite Cells}

In our opinion there can be little doubt that phase-delayed recombination ($A_R$) is responsible for the giant photo-induced capacitances observed in perovskite solar cells, since the only mechanism required is a slow-moving agent acting on carrier densities and thereby recombination rates. Studies of I-V hysteresis have already amply demonstrated that mobile ions play just such a role in the halide perovskites, and our models constitute a further demonstration that the theory works quantitatively in the frequency-domain. Given that similar measurements have been the subject of considerable interest and controversy in the literature however, we feel it is important to spend some time in the following to consider alternative explanations. These other theories are primarily based on attributions to the other half of the measured capacitance, i.e. $C_Q$ instead of $A_R$.

The observation of a large increase in low-frequency capacitance under illumination was first reported in the guise of a giant AC dielectric constant (an experimentally derived quantity simply proportional to $C$) \cite{Juarez-Perez_Photoinduced_2014}. Initially this observation was considered as evidence of light-induced changes to the fundamental dielectric constant, i.e. to the electronic or ionic polarizability, a notion that was greeted with both widespread interest \cite{Grtzel_The_2014,chen2014emergence,yang2015significance,Jung_Perovskite_2015} and skepticism \cite{zhang2015charge,Tress_Understanding_2015,Almond_An_2015,Frost_What_2016}. A closely related observation is the commonly reported growth of AC conductivity with light intensity \cite{Miyano_Simple_2015,Pockett_Characterization_2015,Zaraza_Surface_2016,Pockett_Microseconds_2017}, which manifests as a shrinking of the impedance in Nyquist plots. Without invoking changes to the fundamental dielectric constant, increases in the AC capacitance and conductivity can be simply understood as a consequence of larger recombination rates occurring under illumination \cite{Almond_An_2015,Frost_What_2016}, which in turn contribute to both the real and imaginary parts of the admittance as in equation (\ref{Ysplit}). Increased recombination primarily affects the real part (conductivity), with contributions to the imaginary part (capacitance) also appearing at lower frequencies due to ion-induced delay. A similar explanation was offered soon after the first report of this phenomenon \cite{Almond_An_2015}, but before the widespread recognition that ionic responses are fundamental in the transient response of perovskite cells. Disordered conduction pathways incorporating capacitive gaps were therefore proposed as a mechanism responsible for causing delay, the importance of which remains unestablished. Another point in favour of the ionic theory advanced here is that recombination rates have a direct effect on $A_R$, with higher rates straightforwardly leading to larger values of the low-frequency illuminated capacitance. Several studies have indeed observed a correlation between large values of the illuminated low-frequency capacitance, the degree of hysteresis and poor cell performance \cite{Kim_Control_2015,Elumalai_Hysteresis_2016}, which is entirely natural when seen in this light.

Quantitatively any theory of the photo-induced low-frequency capacitance in terms of accumulated carriers or ions (i.e. any theory invoking $C_Q$ in the language of this paper) is problematic. If the capacitance were a result of photogenerated charges separated across an interface, as in the theories of refs. \cite{Zarazua_Light_2016,Almora_Do_2018}, the observed values of \SI{e-1}{ \farad \per \centi \metre \squared} (Fig. \ref{Fig1}a) would imply an average separation $d=\epsilon \epsilon_0/C$ less than \SI{1}{\pico \metre} between compensating charges. This is clearly unphysical as a length-scale for quantum confinement, being vastly smaller than a typical electron de Broglie wavelength or orbital size. A further problem with the theory of ref. \cite{Zarazua_Light_2016} is that a realistic treatment of accumulation capacitance must also take account of Fermi-Dirac statistics in order to reach the relevant limit of strong accumulation (Fermi level deep in the valence band), the neglect of which leads to inflated estimates. Including only a generous \SI{0.3}{\nano \metre}  distance of closest approach (approximately one bond length) already limits the accumulation capacitance to being less than \SI{2e-4}{\micro \farad \per \centi \metre \squared}, aside from limits due to Pauli exclusion. Removing the restriction that the compensating charge should be separated across an interface brings in the bulk chemical capacitance of electrons and holes as a possible explanation \cite{Sah}, but here extreme carrier densities of order $\SI{1e21}{\per \centi \metre \cubed}$ across the whole device would be required to explain figures of order \SI{e-1}{ \farad \per \centi \metre \squared}. The raw accumulation of carriers or ions must therefore be deemed an insufficient mechanism for explaining the photo-induced low-frequency capacitance. The reported evidence in favour of such theories \cite{Gottesman_Dynamic_2016,Bergmann_Local_2016,Contreras_Specific_2016} includes an approximate linear scaling in the capacitance with light-intensity, and increasing values with larger perovskite thickness \cite{Zarazua_Light_2016}.  These features also follow naturally if the capacitance stems from phase-delayed recombination currents as argued here: since increasing the light intensity induces higher carrier densities in the bulk and concomitantly larger recombination currents. For essentially the same reason,  increasing the film thickness will also enlarge the total recombination current, either because larger carrier densities are required to carry the current, or because the diffusion length is exceeded. Both of these variations in light intensity and film thickness will therefore influence the component of phase-delayed recombination ($A_R$) and hence the measured capacitance. The exponential dependence on open-circuit voltage with a logarithmic slope of $1/2 k_B T$, touted as a unique characteristic of the accumulation capacitance \cite{Zarazua_Light_2016}, is simply indicative of the usual relation between the light intensity and open-circuit voltage $I_{L} \sim \exp( V_{oc}/\gamma k_B T)$ (with an ideality factor $\gamma=2$), and therefore follows from the just established relation between light intensity, recombination current and $A_R$ contributions to the measured capacitance (the result being that $A_R \propto I_L \propto \exp( V_{oc}/\gamma k_B T)$). 

Photo-enhancement of the capacitance can also be observed in the high-frequency region. In the C-f measurement of Fig. \ref{Fig1}a, an excess of approximately \SI{300}{\nano \farad \per \centi \metre \squared} manifests between \SI{1}{\kilo \hertz} and \SI{10}{\kilo \hertz} under illumination compared to the dark measurement. Simultaneously, an additional intermediate arc appears in the Nyquist plot (Fig. \ref{Fig1}a inset). This feature was seen in several batches of cells, and appeared variously as either a distinguished arc in the Nyquist plot or simply as a broadening of the high-frequency feature. A similar feature has been reported previously in cells with a (poorly performing) Nb$_2$O$_5$ electron transport layer (ETL) \cite{Guerrero_Properties_2016}. The question arises as to whether the light-induced capacitance in this frequency range is attributable to $C_Q$, due to the accumulation of poorly extracted carriers or perhaps a photo-doping effect, or alternatively to $A_R$. Although far from being conclusive confirmation, such a feature is already present in the simulation of Fig. \ref{Fig1}b (note the excess $A_R$ capacitance above \SI{1}{\kilo \hertz} which exceeds the capacitance simulated without illumination, denoted by dotted lines). In these simulations the onset of the light-induced high-frequency feature signals an accumulation of carriers at the perovskite/titania interface. Despite contributing only a small amount of stored charge, the delay induced by the charging process causes part of the recombination current to fall out of phase, resulting in significant contributions to the AC capacitance through $A_R$. Alternative theories for the light-induced high-frequency capacitance seen in Fig. \ref{Fig1}a can also be imagined, and may justify further study. However, the resemblance in these simulations is an important indication that phase-delayed recombination currents can also be significant beyond the ionic cutoff (extrapolated to approximately \SI{1}{\kilo \hertz} in Fig. \ref{Fig1}a) where purely carrier dynamics are at play.

In the absence of illumination, recombination currents are very small, particularly at modest bias voltages. For this reason the un-illuminated measurement in Fig \ref{Fig1}a, which features a low-frequency 0V capacitance of \SIrange{1}{10}{\micro \farad \per \centi \metre \squared} at around \SI{1}{\milli \hertz}, is almost certainly a reflection of the $C_Q$ capacitance associated with interfacial ion accumulation. A realistic theory of this capacitance is likely to be complex for a number of reasons. One is that the interfaces in a standard perovskite cell known to be imperfectly blocking to ions: in particular, iodine has been observed to diffuse into the Spiro-OMeTAD contact under operating conditions \cite{jeangros2016situ}. This means that at least one interface is capable of storing large amounts of ionic charge, but more is required to explain the measurements of Fig. \ref{Fig1}a since both perovskite interfaces contribute in series to the measured capacitance. A large measured capacitance therefore implies significant charge-storage ability at both interfaces. We are not aware of any direct evidence indicating that TiO$_2$ is permeable to ions. How then can a large capacitance at the TiO$_2$ interface be explained? This question is explored further in the supplementary information: for our present purposes it is sufficient to note that corrections to the standard IDD model were necessary to achieve dark capacitances above $1$\SI{}{\micro \farad \per \centi \metre \squared} in the simulation of Fig. \ref{Fig1}b. These were made via the addition of surface states at the perovskite/contact-layer interfaces, which allow for the accumulation of larger ionic space charges than would be possible otherwise without using unrealistically large contact-layer doping, although we emphasize that such corrections are in no way crucial to the effect of giant illuminated $A_R$ capacitance discussed above. Because the capacity of a particular interface to accommodate ionic charge is highly dependent on its specific properties, such as permeability, interfacial defect densities and doping, we refer to the dark capacitance in Fig. \ref{Fig1}a as the ``ion-electrode'' capacitance, denoted as $C_{IE}$ in the following. This capacitance determines the timescale of electrode polarization (i.e. the rate at which bulk electric fields are screened by interfacial ion accumulation),  naively as $R_{ion} \cdot C_{IE}$ with $R_{ion}$ the bulk ionic resistance. The implied dependence on electrode properties such as doping, dielectric constant and thickness may explain many of the observed differences in hysteresis between normal and inverted structure cells, which differ primarily in their selective contact layers \cite{Kim_Control_2015,Elumalai_Hysteresis_2016,Liu_Effects_2017}.

\begin{figure*}[t!h!]
	{\centering
		\includegraphics[width=7.0in]{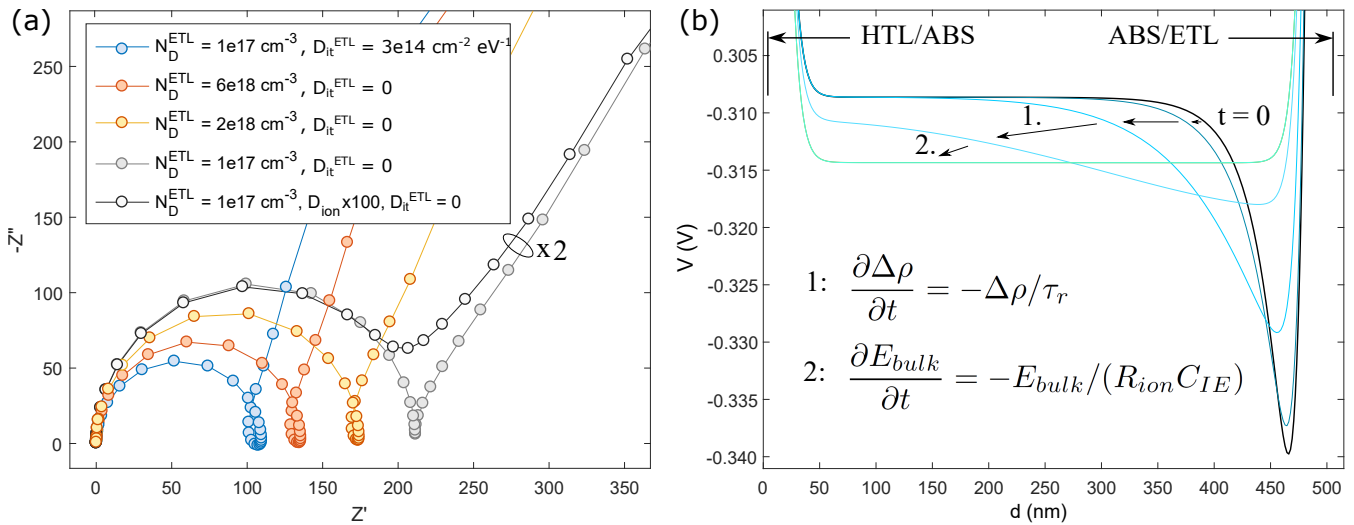}
	}
	\caption{(a) Simulations exhibiting a loop feature at 0.6V under illumination. The data appearing in grey and black have been scaled down by a factor of 2 for visibility. ETL properties that affect the high-frequency capacitance  are found to have an important impact on the loop feature (here the doping $N_D$ and interfacial defect density $D_{it}$). Loops can also be obscured by the high frequency feature if the ions are sufficiently fast (black circles). (b) Time domain simulation of the model in (a) (with $\mathrm{N_D^{ETL}}=\SI{6e18}{\per \centi \metre \cubed}$) showing the evolution of the bulk potential (referenced to the grounded ETL-side metal contact) in response to a step change in applied potential (\SI{0.6}{\volt}$\rightarrow$\SI{0.8}{\volt}). Contact layers (HTL and ETL) are located at $x=0$ and $x=\SI{500}{\nano \metre}$ respectively. Ionic relaxation occurs on the timescale $\tau_r$ (step 1) and interfacial charging on the ionic $RC$ timescale (step 2) according to the labelled equations.}
	\label{Fig2}
\end{figure*}

\section*{Loop Features and Negative Capacitance}
Distinct features occasionally appear in EIS measurements at intermediate frequencies (\SI{1}{\hertz}-\SI{1}{\kilo \hertz}), in addition to the usual high and low frequency responses \cite{Pockett_Microseconds_2017}. The most intriguing intermediate-frequency feature manifests as a loop in the complex impedance plot \cite{Guerrero_Properties_2016}, sometimes extending below the real axis signifying a negative value of the AC capacitance. Suggestions for the meaning of these features include the interplay of an unidentified intermediate state \cite{Guerrero_Properties_2016}, an interfacial charge transfer resistance \cite{Ghahremanirad_Inductive_2017}, and genuine inductance (i.e. energy storage in magnetic fields) associated with the ionic conduction current \cite{Kovalenko_Ionic_2017}. Whilst the former two remain plausible, if vague, the use of inductors in equivalent circuits should not be conflated with physical inductance \cite{Tada_Comment_2017} and is simply a mechanism for modelling negative capacitances (actual negative capacitor elements would be no less physical). Instead, the emergence of loop features in our IDD simulations suggests that loop features are related to $A_R$ on the timescale of ionic relaxation:
\begin{equation} \tau_r = \frac{\epsilon \epsilon_0}{\sigma_{ion}} . \label{taur} \end{equation}
where $\epsilon$ is the perovskite's relative permittivity and $\sigma_{ion}$ is the bulk ionic conductivity. In the theory of conventional semiconductors this quantity (expressed instead in terms of the carrier conductivity) is the timescale at which charge density perturbations are screened by the majority carrier \cite{Roosbroek}. The relaxation time $\tau_r$ is therefore expected to play some role whenever changes in the applied potential induce charge density variations at times $t<\tau_r$ in the bulk perovskite. Note that $\tau_r$ refers to screening by ionic charges in the bulk, and therefore differs in general from the timescale of electrode polarization, which depends on several properties of the electrodes and the perovskite thickness (absent from $\tau_r$). The most natural way for $\tau_r$ to become involved is if sizeable carrier densities are present in the bulk of the perovskite layer: in such cases, carrier screening of the applied potential will result in bulk charge density variations at (very short) times that will be subsequently screened by ions on the timescale $\tau_r$. Since the standard IDD models with perfectly compensated ions are essentially free of bulk carriers for low to moderate bias voltages, some modification is needed to introduce a significant carrier density and thereby make the relaxation time manifest.

We note in advance of the following that if the above theory is correct, loop features should allow identification of the ionic conductivity through equation (\ref{taur}). In principle this can then be used to estimate the ion density, although presently a large range of ionic diffusion constants (\SI{1e-8}{\centi \metre \squared \per \second} to \SI{1e-12}{\centi \metre \squared \per \second} for the iodide vacancy in MAPbI$_3$) \cite{Yang_The_2015,Eames_Ionic_2015} hinders the accuracy of this step. Observations of a loop feature at approximately \SI{1}{\kilo \hertz} \cite{Guerrero_Properties_2016} in this sense suggest ion densities between \SIrange{1e17}{1e21}{\per \cubic \centi \metre \cubed}, which is at least in approximately the expected range \cite{Walsh_Self_2015}.

%
Fig \ref{Fig2}a shows a simulated impedance spectrum featuring an ``inductive loop'' at intermediate frequencies, produced by including a net acceptor doping density at a significant fraction of the mean ion density ($N_A = \SI{4e17}{\per \centi \metre \cubed}$ with $N_{ion}=\SI{1e18}{\per \centi \metre \cubed}$). This net doping produces the required excess carrier density motivated above. Figure \ref{Fig2}a also shows a few variations of the base model with respect to the contact layers properties that we will return to shortly. In this model we used a slightly lower ion density than the default of \SI{1e19}{\per \cubic \centi \meter} to ensure that the doping density was not fully compensated by ion migration from the interfaces (see supplementary Fig. S2). Since the primary requirement is a large bulk carrier density, an alternative method of engineering the loop feature would be to include an unfavourable band offset at one of the contact layers, such that photogenerated carriers of one species or the other are accumulated in the absorber. Regardless of the exact mechanism, the loop in such models is associated with a negative contribution to the total capacitance by phase-delayed recombination ($A_R$), and extends below the real axis only when this contribution overwhelms the remaining positive capacitance in $C_Q$. 

The reason why ionic screening at the relaxation time manifests as negative $A_R$, and therefore a loop feature, is best understood in the time domain. Figure \ref{Fig2}b plots the electrical potential across the bulk of the perovskite, showing the time evolution following a step change in applied potential. It is seen that the system responds at short times ($t<\SI{1}{\micro \second}$) with the development of an electric field at the perovskite/titania p-n junction ($t=0$ in the figure). This initially localized potential drop is caused by a rearrangement of the majority holes within the depletion region, which entails a change in the bulk charge density. This perturbation is therefore screened away to the interfaces by ionic relaxation on the timescale $\tau_r$ (labelled as step 1 in the figure), leaving an approximately uniform electric field (linear potential drop) across the bulk. Eventually this uniform electric field is fully screened from the bulk by an exchange of ionic charge between the interfaces on the slower (RC) timescale of electrode polarization (step 2). The intermediate process of screening an inhomogeneous field into a uniform one temporarily increases recombination (i.e. results in a transient characterized by $R_{\tau}<0$, c.f. equation \ref{Rt}), in this model of the minority electrons at the HTL interface. We do not claim surface recombination plays a unique role however, as other recombination models could plausibly behave in a similar fashion. Increased recombination on the timescale $\tau_r$ results in a negative contribution to the capacitance as per equations (\ref{A_R}) and (\ref{Cdebye}), with positive capacitance taking over again at low frequencies (the timescale of electrode polarization) as in Fig. \ref{Fig1}.

Intermediate loop features have been reported by a few groups in the literature \cite{Guerrero_Properties_2016,Wang_Insights_2017,Kovalenko_Ionic_2017}, but were not encountered in any of the cells handled for this study. In simulations the phenomenon is found to be somewhat delicate, being easily obscured by the standard low and high-frequency features. We infer from numerical experiments that a distance of some 1-2 decades between $\tau_r$ and the other high and low-frequency $RC$ time-constants may be necessary for loop features to emerge. The intermediate relaxation time $\tau_r$ expressed as a fraction of the low frequency $RC$ time-constant (assuming $R=\rho_{ion}L$) is simply $C_{geo}/C_{IE}$, the ratio of the geometric to ion-electrode capacitance. Simulations with a small ion-electrode capacitance consistently fail to produce loop features, likely because the ratio $C_{geo}/C_{IE}$ is not small enough to ensure a separation of the loop (or negative capacitance contribtion) from the low-frequency feature. In IDD models the ion-electrode capacitance is determined by the ion density (which contributes a diffuse Debye-layer capacitance) and electrode properties depending on how the ions are compensated (e.g. doping density if by exposed dopants, or surface state density if by trapped carriers). This results in a strong dependency of the loop features on electrode properties as seen in Fig. \ref{Fig2}a. Loops can also be obscured if the ions are fast enough for $\tau_r$ to run into the high frequency feature (Fig \ref{Fig2}a, black markers). These considerations may explain some of the observed sensitivity of loop features on properties of the titania layer \cite{Guerrero_Properties_2016,Anaya_Electron_2016}. Another consideration is that the contribution of $A_R$ will only be appreciable if a significant portion of the applied potential is dropped within the perovskite layer where ionic screening can have an effect. For models with depleted, thick titania layers and no surface states this requirement is not met, since most of the potential drops within the titania, explaining why the loops are not visible in these simulations (Fig \ref{Fig2}a, grey markers). Enlarging the doping density (red and yellow markers) or adding surface states (blue markers) serves the dual purposes of increasing the ion-electrode capacitance and also increasing the amount of applied potential dropped within the perovskite layer, both of which lead to more significant loops in simulation. Almost all cell parameters have an effect on $A_R$ however, so that isolating the experimentally relevant factors wherever loops appear will remain a challenging task. 

\begin{figure}[h!!]
	\includegraphics[width=3.4in]{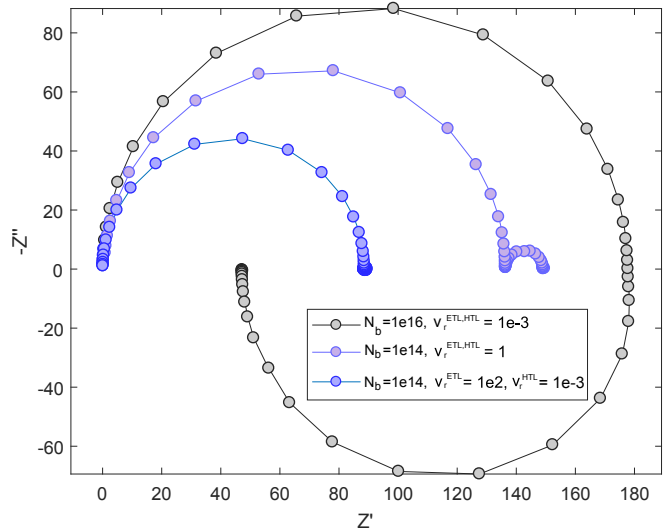}
	\caption{ Simulated Nyquist plots (\SI{1}{\milli \hertz} to \SI{1}{\mega \hertz}) under dark conditions and 1V forward bias for differing recombination models characterized by a single bulk SRH defect (density $N_b$, expressed in \SI{}{\per \centi \metre \cubed}) and surface recombination (velocity $v_r$, expressed in \SI{}{\centi \metre \per \second}) at each perovskite interface (see supplementary table S1 for further details). Here a negative low-frequency capacitance is observed in a model with a high density of bulk recombination centres (black series). }
	\label{Fig3}
\end{figure}

Negative capacitance in the low-frequency spectrum has also been reported several times, both under illumination \cite{Sanchez_Slow_2014,Anaya_Electron_2016,Fabregat-Santiago_Deleterious_2017} and in darkness \cite{Dualeh_Impedance_2014,Sanchez_Slow_2014,Zohar_Impedance_2016}. In either case its physical origin has not been conclusively identified. Often, the low-frequency negative capacitance emerges under conditions where injected current is either dominant (under dark conditions) or a significant contributor (at large forward bias under illumination) to the total current. This raises the possibility that modulation of the injected current, rather than of the collected photocurrent, could be responsible for negative capacitance contributions through $A_R$. It has been shown previously that whilst ion accumulation below the built-in voltage generally increases the recombination of photogenerated carriers, it also acts to retard carrier injection,  or equivalently the bulk recombination current under dark conditions   \cite{Jacobs_Hysteresis_2017}. Said otherwise, ionic redistribution in IDD models frequently has opposite effects on the recombination rate of photogenerated carriers ($R_{\tau}>0$, see discussion above) and that of injected charge carriers ($R_{\tau}<0$). Negative contributions to the capacitance from $A_R$ may  therefore be anticipated whenever the injected current is a significant fraction of the total current. According to our numerical experiments, some of which are shown in Fig. \ref{Fig3}, this should not be regarded as a rule but simply as a possibility, since the magnitude and sign of the low-frequency capacitance is found to depend on the recombination model. In the simulations of Fig. \ref{Fig3}a a bulk recombination model resulted in negative capacitance (black markers), whereas surface recombination models showed  positive capacitance with less overall sensitivity to the electrode polarization at low frequencies (blue markers). Similar behaviour can be expected under illumination if the bias voltage is large enough to produce a significant injected current. The suggestion that negative low-frequency capacitance is a detrimental indication for cell performance \cite{Fabregat-Santiago_Deleterious_2017} may then be justified if these cells are ``leaky'', i.e. have a particularly high dark current (comparable to the collected photocurrent under illumination and forward bias). Alternatively, we note that in time-domain I-V measurements the unusual situation $R_{\tau}<0$ (corresponding to negative capacitance) is referred to as ``inverted-hysteresis'' \cite{Tress_Inverted_2016}, and is interpreted in terms of ion-modulated collection efficiency (equivalently recombination). Typically inverted hysteresis is also only observed at forward bias voltages above the cell's built-in voltage, and is likely explained by ion-enhanced surface recombination \cite{Shen_Inverted_2017}.

\section*{The high-frequency capacitance: C-V measurements}

\begin{figure*}[t!]
	\includegraphics[width=7in]{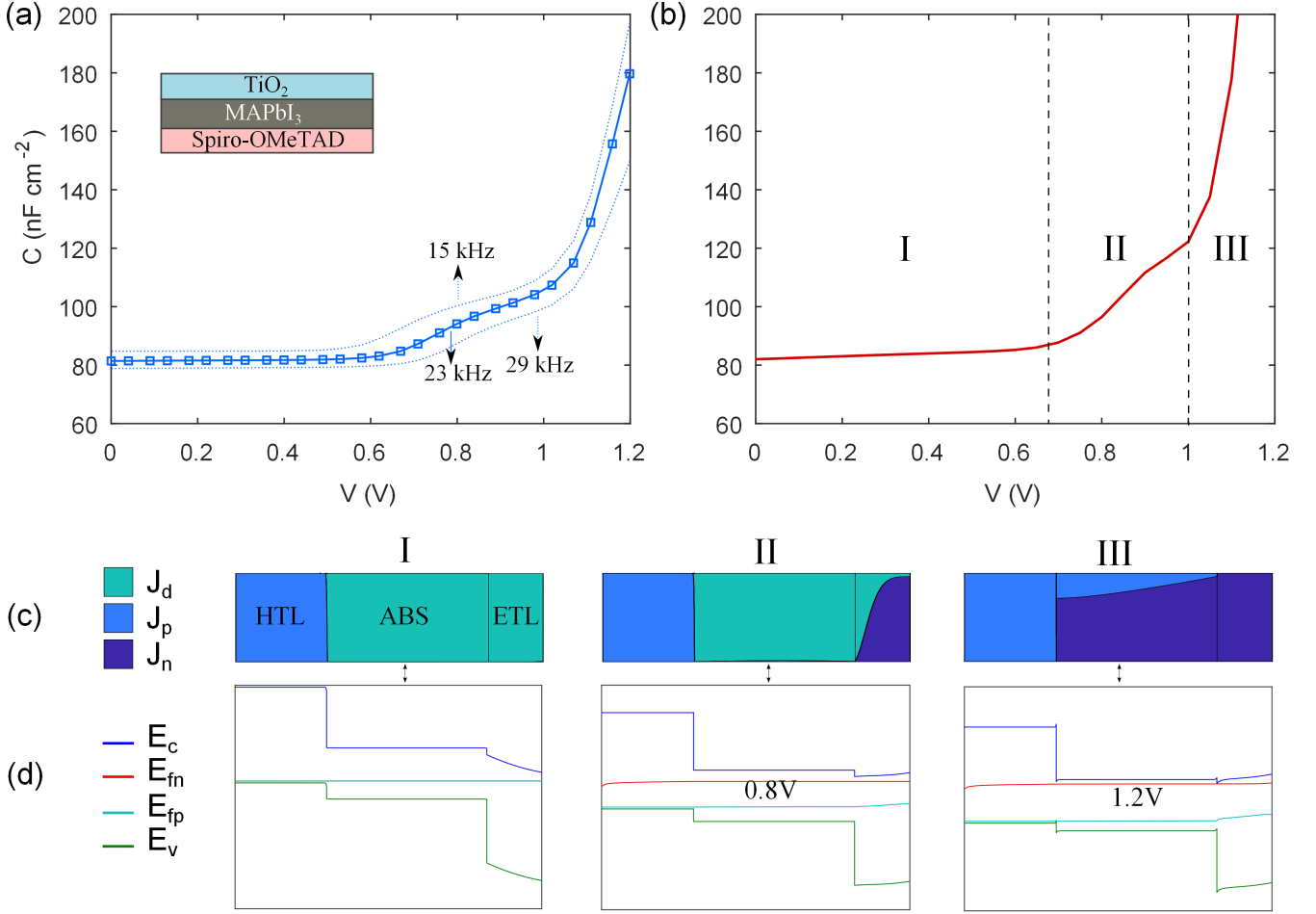}
	\caption{(a) Slow-scan measurements of the un-illuminated capacitance versus voltage taken on a standard MAPbI$_3$ perovskite cell. (b) Simulations of the same measurement (parameters in table S1). (c) Cumulative area plots of the AC current taken from the simulation in (b) showing spatially resolved contributions due to electron and hole conduction ($J_n,J_p$), as well as displacement ($J_d$). (d) Band diagrams corresponding to (c), illustrating depletion in the ETL in region I and accumulation in region III. Flat bands are present in the perovskite region due to ionic screening of the built-in potential.}
	\label{Fig4}
\end{figure*}

In the above we have shown that the phase-delayed recombination $A_R$ plays a significant role in capacitance measurements under illumination, and also in darkness at low-frequencies and high bias (both conditions where the recombination current is significant compared to the charging current). By contrast, high-frequency C-V measurements taken under dark conditions provide a case where the charging current $Y_Q$, and therefore $C_Q$, can be expected to dominate the measured capacitance. Given the success of the IDD models so far in accounting for the ionic influence on recombination currents, it is of interest to know whether these models adequately represent the charging of perovskite cells as well. In the following we evaluate our EIS simulations against high-frequency C-V measurements taken on TiO$_2$/MAPbI$_3$/Spiro-OMeTAD cells.

An immediate issue with C-V measurements of perovskite cells is that, in common with I-V measurements, there is often clear rate-dependent hysteresis in their C-V sweeps. For simplicity here we focus on slow scans approximating a steady-state measurement, although in our experience the amount of hysteresis observed at rates as low as 0.1mV/s was often still considerable. In the measurement of Fig. \ref{Fig4}a a slow scan rate of  0.7mV/s was used. Plots of the hysteresis in these measurements are included in supplementary Fig. S1. At each voltage step a frequency scan was performed in the range of 1kHz-1MHz, revealing a small degree of frequency dependence which will be ignored hereafter (this is indicated by the dotted lines in Fig. \ref{Fig4}a, and is possibly caused by microscopic inhomogeneities due to the mesoporous TiO$_2$ scaffold \cite{Almond_An_2015}). Simulations of this measurement were found which strongly resemble the measurement as shown in Fig. \ref{Fig4}b; the adjusted but physically reasonable parameters used are listed in table S1. For this simulation the ions were relaxed for \SI{1000}{\second} at each bias point: relaxing for a shorter period of time reveals that these models also qualitatively reproduce the C-V hysteresis observed in forward-reverse scan measurements (Figs. S1a,b,d). In the following we will first provide an interpretation of the simulation results before assessing their correspondence with the measurements in Fig. \ref{Fig4}a. 

Both the measured and simulated data in Fig. \ref{Fig4} can be split into three voltage regions according to their qualitative behaviour as done in Fig. \ref{Fig4}b. Plots of the total (AC) current, spatially resolved into components of electron current, hole current, and displacement current in Fig. \ref{Fig4}c serve to illustrate the mechanisms at play in each region. We note that the possibility of drawing such figures is guaranteed by the fact that AC current, including the displacement contribution, is spatially constant across a 1-dimensional device. According to these figures, region I is defined by current flowing as displacement through both the perovskite and titania layers, making the capacitance in this voltage range relatively small (a serial combination of the individual geometric or parallel-plate capacitances of each layer). In region II the application of forward bias replenishes the depleted titania layer, shorting out its serial contribution to the capacitance and leaving only the perovskite layer as dielectric spacer. Finally in region III, whose onset is approximately the built-in voltage, the accumulation of carriers within the perovskite layer causes a rapid increase in the capacitance with voltage (a combination of $C_Q$ and $A_R$ components, with the former dominating in this case).


The apparent agreement between measurement and simulation in Fig. \ref{Fig4} is striking but requires further scrutiny. In our simulations, full depletion of the titania layer (as seen in the band-diagram of Fig. \ref{Fig4}d, left) occurs due to ionic redistribution, and accounts for the relatively constant capacitance in region I of the simulations. This is a result of the tendency for mbile ions to screen the built-in voltage away from the bulk of the perovskite, causing it to drop both inside the contact layers and in the narrow diffuse ionic layers adjoining them (the latter are invisibly small in Fig. \ref{Fig4}d). The specific parameters used in the simulation of Fig. \ref{Fig4}b  include contact layer doping densities of \SI{1e19}{\per \centi \metre \cubed} for the HTL (``Spiro-OMeTAD'') and \SI{1e17}{\per \centi \metre \cubed} for the ETL (``titania''), in addition to high ion densities \SI{1e19}{\per \centi \metre \cubed} within the perovskite. These parameters cause most of the built-in voltage to drop within the titania due to its lower doping, leaving it fully depleted at $0$V  (see Fig. \ref{Fig4}d). Our choice of a high doping density for the Spiro layer is motivated by Kelvin Probe Force Microscopy (KPFM) studies which indicate that very little of the built-in and applied voltage drops within the Spiro layer \cite{Jiang_Carrier_2015,Guerrero_Electrical_2014}, suggestive of small depletion widths. However, as mentioned previously, this could alternatively be a result of ion uptake from the perovskite layer instead of intentional doping, a process that has been clearly observed for several intrinsic and extrinsic species \cite{Zhao_Mobile_2017,Carrillo_Ionic_2016,Ginting_Degradation_2017,Li_Extrinsic_2017,jeangros2016situ}.  Setting larger doping in the titania layer causes the extent of region II to reach downward to lower voltages, eventually entering the negative bias range. Examples of simulations with a larger titania doping are provided in the supplementary information (these more closely resemble some previously reported C-V  measurements \cite{Almora_On_2016}). Since the doping of compact and mesoporous titania layers can vary widely with deposition and annealing methods, it is not surprising that previous reports have differed in their evaluation of whether the depletion width in titania is significant \cite{Guerrero_Electrical_2014,Liu_Electrical_2014,Laban_Depleted_2013}. Indeed, several studies have interpreted the capacitance seen in region II as being due to carrier accumulation in the perovskite layer (a result of p-type doping) \cite{Liu_Electrical_2014}, disregarding the effect of titania depletion by comparison \cite{Almora_On_2016}.

\begin{table*}[t!]
	\begin{center} 
		
		\begin{tabular}{| m{5cm} | m{6cm} | m{6cm} | }
			\hline			
			 \centering Feature &  \centering Interpretation (This Work) & \begin{center} Interpretation (Literature)  \end{center} \\ \hline
			 \centering Photo-induced Low-frequency Capacitance (light) & \centering $A_R$: ionically modulated recombination (timescale $R_{ion} C_{IE}$) & $C_Q$: light-induced polarizability and/or polaron hopping \cite{Juarez-Perez_Photoinduced_2014}, carrier plus ion accumulation \cite{Zarazua_Light_2016,Almora_Do_2018}, light-induced ion migration \cite{bag2015kinetics}. \par $A_R$: delay caused by random microstructure with capacitive gaps \cite{Almond_An_2015}. \\ \hline
			\centering Inductive loop features (light) & \centering $A_R$: ionically modulated recombination (timescale $\tau_r$) & Interaction of an unspecified intermediate state \cite{Guerrero_Properties_2016}, interfacial transfer resistance \cite{Ghahremanirad_Inductive_2017}, physical inductance \cite{Kovalenko_Ionic_2017}.  \\ \hline
			\centering High-frequency Capacitance (dark) & \begin{itemize} \item $V < \SI{0.7}{\volt}$: serial combination of the perovskite and titania geometric capacitance ($C_Q$). \item $\SI{0.7}{\volt} < V < \SI{1.0}{\volt}$:  serial combination of the perovskite geometric and titania depletion capacitance ($C_Q$).  \item $V>\SI{1}{\volt}$: carrier accumulation plus recombination ($C_Q+A_R$). \end{itemize} \begin{center} Voltage ranges are approximate. \end{center} & \begin{itemize} \item $V < \SI{0.7}{\volt}$: Perovskite geometric capacitance \cite{Almora_On_2016}, double-layer capacitance \cite{Pascoe2015}, carrier accumulation \cite{Kim_Mechanism_2013}. \item $\SI{0.7}{\volt} < V < \SI{1.0}{\volt}$:  depletion capacitance of a doped perovskite layer \cite{Almora_On_2016,aharon2014depletion}, double-layer capacitance \cite{Pascoe2015}, carrier accumulation \cite{Kim_Mechanism_2013}.  \item $V>\SI{1}{\volt}$: carrier and/or ion accumulation \cite{Almora_On_2016,Kim_Mechanism_2013}. \end{itemize} 
			\begin{center} Voltage ranges are approximate. \end{center} \\ \hline
			\centering High-frequency Capacitance (light) & \centering As above with small contributions from carrier accumulation and/or $A_R$ as observed in the simulation of Fig. \ref{Fig1}b & \begin{center} As above. \end{center} \\ \hline
			\centering Low-frequency Capacitance (dark) & The ion-electrode capacitance $C_{IE}$ plus ionically modulated charge injection and recombination. The $A_R$ component will grow approximately exponentially with applied voltage and can take either sign. & \begin{center} Deep-level defects \cite{Samiee_Defect_2014,Duan_The_2015,shao2014origin,dong2015electron,lee2015formamidinium}, dielectric relaxation \cite{bisquert2014theory,Sanchez_Slow_2014,Pascoe2015}, electrochemical reaction (negative capacitance) \cite{Zohar_Impedance_2016}. \end{center}  \\
			\hline   
		\end{tabular}
	\end{center}
\caption{A summary of what we consider to be the primary contributions to features seen in typical capacitance measurements of perovskite solar cells, compared to a sampling of conflicting interpretations from the literature (this is not a comprehensive list). Where appropriate interpretations are labelled in parantheses according to whether they invoke primairly $C_Q$, $A_R$ or both, as defined by equation (\ref{A_R}).}
\end{table*}

To verify our prediction of depletion in the titania layer experimentally, otherwise identical MAPbI$_3$ cells were prepared on titania/FTO substrates with varying compact layer thicknesses (by spinning 2, 4 and 6 layers according to procedures published elsewhere \cite{Shen_Improved_2017}). It was found that the low-voltage capacitance (region I) scales inversely with the titania thickness (Figs. S1a-c), strongly indicating that the depletion width in titania extends over the full extent of the compact layer. If the depletion width was only a finite fraction of the titania thickness, then no dependence of the capacitance in region I on the titania thickness would be expected. Furthermore, no correlation was found between the capacitance in regions II or III and the titania thickness (see Figs. S1a-b). It therefore seems clear that the capacitance in region I of our measurements in Figure \ref{Fig4}a includes the geometric capacitance of a largely depleted titania layer. Whether the perovskite remains essentially free of carriers in region II as suggested by our simulations, or whether it also contributes to the depletion capacitance, is less certain. On this issue it must be noted that the concept of a macroscopically p-type (or n-type) perovskite layer becomes more complex in the IDD theory because ionized dopants can be compensated by mobile ions (which act themselves as a mobile doping density), instead of, or in addition to excess carriers \cite{Walsh_Self_2015}.  As we show in the supplementary information (Fig. S2), the IDD theory predicts that a large compensated ion density (order \SI{1e19}{\per \centi \metre \cubed}) will readily screen any excess charge away from the bulk up to a comparable density, bringing in far fewer than the expected number of carriers for compensation. For this reason the standard picture of a p-n type junction between perovskite and titania \cite{Guerrero_Electrical_2014} is not comfortably compatible with the IDD model and a large density of compensated ions. This issue is entirely averted if the perovskite remains essentially intrinsic in terms of carrier density and depletion is located in the titania instead, where a static and approximately uniform doping profile of order \SI{1e17}{\per \centi \metre \cubed} is within the expected range \cite{ORegan_Vectorial_1990}. We conclude that the C-V measurements in Fig. \ref{Fig4} can be understood using the IDD model if the depletion capacitance in region II is assigned to the titania layer instead of the perovskite. This does raise the problem of how to account for the negative space charge that appears in KPFM measurements as part of a p-n junction between the perovskite and titania, extending over as much as 300nm into the bulk \cite{Guerrero_Electrical_2014,Jiang_Carrier_2015}. If these are an accurate representation of bulk electrical field profiles then ion densities must be much lower than density functional theory calculations would indicate \cite{Walsh_Self_2015}, since the predicted densities yield Debye lengths on the order of a few nanometers. In any case, Mott-Schottky analysis as applied to the perovskite layer is unlikely to yield reliable results as it assumes a static charge distribution compensated solely by carriers, leaving ions unaccounted for.

\section*{Summary of interpretations}
The previous sections have built up a set of  interpretations for features commonly seen in the EIS spectra of perovskite cells based on our drift-diffusion device models. These are summarized in table I where we also list a number of alternative hypothesis put forward in the existing literature, some of which have been addressed already in the relevant sections. We will not attempt to deal with each discrepancy individually, but instead offer three factors which count in favour of the interpretations we are suggesting here. The first is the physical arguments against their alternatives, for example those presented above against $C_Q$-type explanations of the photo-induced low-frequency capacitance. A second factor is the congruence achieved between the numerical simulations presented and our experimental results (Figs. \ref{Fig1}, \ref{Fig4} and Fig. S1). The third and final factor is consistency between simulations of different measurement types, i.e. the fact that all the features discussed in this paper were simulated with a single class of models, being the same models as those used to study I-V hysteresis \cite{Shen_Inverted_2017}. By comparison most prior work has dealt with only one feature or measurement type in isolation. Nonetheless, some of our interpretations lack experimental verification and may require further qualification: this refers in particular to our mechanism for explaining the loop features and the low-frequency negative capacitance. 

Some of our interpretations are contrary to prior work which has applied EIS to determine various device properties. The use of thermal admittance spectroscopy is particularly notable, since this technique has been applied several times to provide supposed measurements of defect densities in MAPbI$_3$ cells \cite{Samiee_Defect_2014,Duan_The_2015,shao2014origin,dong2015electron,lee2015formamidinium}. Deep-level defects are known to produce features in C-f measurements due to the finite timescale associated with capture and emission \cite{Schroder}, and could plausibly constitute part of the low-frequency response seen in perovskite cells. However, given the typical size of the ion-electrode capacitance $C_{IE}$, it would be very challenging to extract the capacitance response due to traps alone, unless this response appeared at higher frequencies beyond the ionic cut-off. To the contrary, the timescale attributed to deep-level trapping in the aforementioned studies coincides with the typical timsecale of ion migration. Furthermore, the expected size of the capacitance contribution from traps is too small to account for typical measurements.  An estimate of the capacitance contribution due to trapping is  $\epsilon \epsilon_0 / \langle x \rangle$, with $\langle x \rangle$ the first moment of the charge density differential (including the compensating charge) caused by trap charging \cite{abou2016advanced}. For trapping in the bulk perovskite this yields an approximate figure of $C\approx \SI{500}{\nano \farad \per \centi \metre \squared}$, assuming a typical length scale of \SI{100}{\nano \metre} for the compensation, i.e. vastly less than the capacitance measured under illumination (Fig. \ref{Fig1}a), but large enough to be at least a contributor to the low-frequency capacitance ($\approx \SI{5}{\micro \farad \per \centi \metre \squared}$) measured under dark conditions. Trapping at the interface layers is a distinct proposition, since such charge can be compensated at short range (of order 1 to \SI{10}{\nano \metre}) by accumulated ions \cite{Richardson_Can_2016}. As discussed above it seems that carrier trapping is a necessary feature, at the TiO$_2$ interface in the least, to explain the large values of $C_{IE}$ seen in dark measurements, since these are not compatible with compensation of the ions across a wide depletion layer (c.f. the discussion of Fig. \ref{Fig4}). 

\begin{figure*}[t!h!!]
	{\centering
		\includegraphics[width=7in]{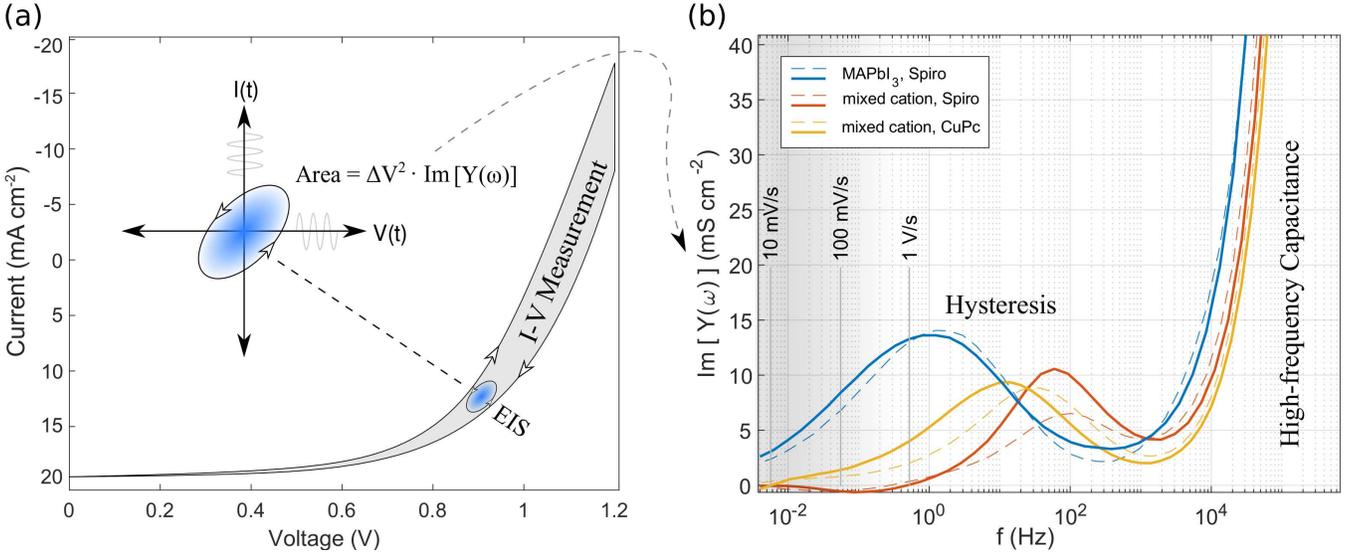}
	}
	\caption{(a) Schematic of the comparison between an I-V sweep and EIS measurement, depicted for a perovskite cell under illumination. (b) EIS measurements of perovskite cells under illumination and biased near their maximum power-point. Plotted is the imaginary part of the admittance, which is simply related to the area of the Lissajous (current-voltage) ellipse as illustrated in (a). Measurements were taken for three cell types based on either the benchmark MAPbI$_3$ perovskite or a mixed cation perovskite, and with either Spiro-OMeTAD or CuPC as the hole-selective contact (full experimental details are given in the supplementary information). Measurements on two distinct cells with each material combination are shown (solid and dotted lines) to indicate the processing variability. The limited frequency range accessed by typical I-V measurements, taken at $10$-$1000$\SI{}{\milli \volt \per \second}, is shaded in grey. The equivalent frequency for I-V measurements is determined via $f= r/ \,2\,(V_{f}-V_{i})$, with $r$ the scan rate and $(V_{f}-V_{i})$ the voltage sweep range.}
	\label{Fig_intro}
\end{figure*}

\section*{Quantifying Hysteresis with EIS}


The theory and simulations presented in this paper provide a bridge between the phenomenology of EIS measurements and the current best understanding I-V hysteresis in perovskite cells. To illustrate the relationship between an EIS measurement and I-V hysteresis we refer to Fig. \ref{Fig_intro}. Figure \ref{Fig_intro}a  schematically depicts a forward-reverse I-V sweep typical of a perovskite cell under illumination: the degree of hysteresis manifested by the discrepancy between these I-V sweeps is regarded as a component of cell performance, as significant hysteresis on slow timescales can interfere with maximum power-point tracking \cite{Pellet_Hill_2017}. Hysteresis is also of interest in fundamental studies which compare cell types characterized by differences in material composition and structure. Furthermore, its evolution with aging is of potential significance in the evaluation of long-term stability, as a large ionic response may coincide with or precede material degradation. It is no surprise then that a large number of recent cell fabrication studies proudly declare the production of cells that are ``hysteresis-free''. The potential pitfall of such an statement is indicated in Fig. \ref{Fig_intro}b, which quantifies the hysteresis measured on three different cell types as determined using EIS. Plotted is the imaginary part of the admittance, or susceptance, which signifies the area of the Lissajous ellipse as illustrated in Fig. \ref{Fig_intro}a. Here it can be seen that although the mixed-cation cells will appear to be practically hysteresis-free devices on the timescale of a typical I-V measurement, say at $100$\SI{}{\milli \volt \per \second} (labelled in the figure), significant hysteresis is nonetheless present, simply at a shorter timescale. This accords with a recent study which derved a fast ion mobility of \SI{3E-7}{\cm \squared \per \volt \per \second} in devices with a mixed cation perovskite \cite{bertoluzzi2018situ}, orders of magnitude faster than has been estimated for MAPbI$_3$ devices. 

As noted elsewhere \cite{habisreutinger2018hysteresis}, the standard method of quantifying hysteresis in terms of indices calculated from I-V data is problematic for a number of reasons. Here we name three which we consider to be the most serious:
\begin{enumerate}
	\item Uniqueness: The I-V sweeps of a cell with significant hysteresis are complex, making it hard to settle on a well-motivated or unique quantifier. Several hysteresis indices have been defined and used in the literature, making comparison across studies difficult.  
	\item Resolution: By sampling the I-V behavior at only a few scan rates one obtains very limited information about the frequency/time-dependence of hysteresis. Incorrect conclusions can be drawn as discussed with reference to Fig. \ref{Fig_intro}b. Although fast I-V scans can be performed with the right equipment, it is generally necessary to re-equilibrate  the device between each sweep to avoid carry-over effects, a time-consuming process if several scan-rates are to be investigated.
	\item Relevance: For practical purposes only the hysteresis behaviour near the operating point (maximum power point under illumination) is of interest for cell performance, making most of the data in an I-V sweep redundant. 
\end{enumerate}

EIS measurements provide a natural alternative to the method of I-V indices. This addresses all three issues above in the following manner:

\begin{enumerate}
	\item Uniqueness: EIS data is completely specified by only two real quantities at each frequency, making it far easier to settle on a unique quantifier. The imaginary component of the admittance is a natural choice. 
	\item Resolution: In EIS it is efficient to gather data over a wide frequency range at high resolution, with only one initial period of stabilization. 
	\item Relevance: By measuring near the maximum power point, at steady-state and under realistic conditions one obtains a measurement of hysteresis without redundancy.
\end{enumerate}

\begin{figure*}[t!]
	\includegraphics[width=7in]{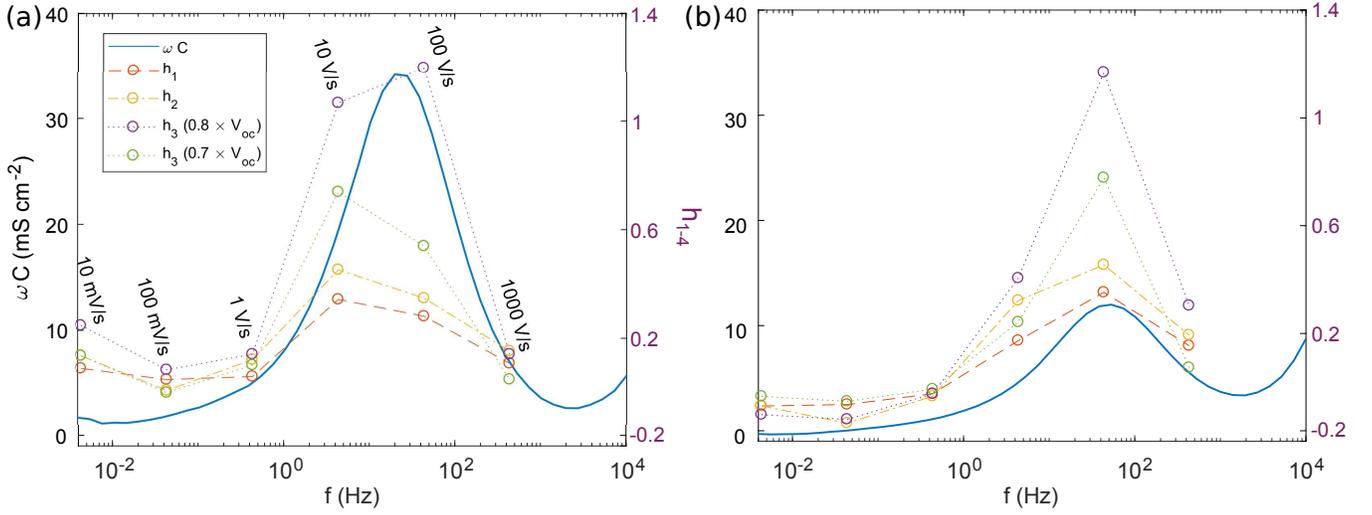}
	\caption{Measurements of hysteresis in terms of $\operatorname{Im}\left[ Y \right] = \omega C$ (susceptance) on two mixed cation cells: (b) with and (a) without fullerene passivation on the TiO$_2$ ETL \cite{peng2017interface}. Each EIS measurement was conducted under approximately 1-sun illumination and near the cell's maximum power point. Shown for comparison are some commonly used hysteresis indices extracted from rate-dependent I-V sweeps measured under the same illumination. The data points correspond to scan rates in the range \SI{10}{\milli \volt \per \second} to \SI{1000}{\volt \per \second}, converted to an equivalent frequency via $f= r/ \,2\,(V_{f}-V_{i})$, with $r$ the scan rate and $(V_{f}-V_{i})$ the voltage sweep range. }
	\label{Fig5}
\end{figure*}

The expression of hysteresis in an EIS measurement is best seen in the Lissajous figure traced by plotting the data as $(V(t),I(t))$. Provided linearity is observed (i.e. the voltage amplitude is not too large), the figure traced is an ellipse with area proportional to the imaginary component of the admittance, or ``susceptance'' as illustrated in Fig. \ref{Fig_intro}a. The area of the Lissajous ellipse $A=\Delta V^2 \operatorname{Im}\left[ Y \right]$, modulo the factor $\Delta V^2$, is a natural way to quantify hysteresis, and coincides in spirit with one commonly used I-V index which measures the area between forward and reverse scans \cite{Levine_Interface_2016}. Since the susceptance is proportional to the AC capacitance ($\operatorname{Im}\left[ Y \right] = \omega C$), plots of $\operatorname{Im}\left[ Y \right]$ also contain the high-frequency capacitance, which is of course another form of hysteresis, albeit one present in any electronic device, and practically irrelevant in the operation of a solar cell. This is the cause of the high-frequency blow-up in Fig. \ref{Fig_intro}b  as $\operatorname{Im}\left[ Y \right] \propto \omega \, C_{hf}$. Ideally one would like to remove this high-frequency component to evaluate only the low-to-intermediate frequency hysteresis induced by ion migration, but this is hardly necessary provided that the features are sufficiently separated. An example of such a ``hysteresis plot'', showing both the hysteresis peaks and high-frequency blow-up, was shown in Fig. \ref{Fig_intro}b; more examples are given in Fig. \ref{Fig5} where only the peaks are shown.

The relationship between the susceptance and several hysteresis indices is indicated by the measurements of Fig. \ref{Fig5}, conducted on two mixed-cation perovskite cells with and without ETL-side fullerene passivation \cite{peng2017interface}. The indices $h_{1-3}$ are three common choices encountered in the literature \cite{Kim_Parameters_2014,Sanchez_Slow_2014,Levine_Interface_2016} defined as  
\begin{align*}
h_1 &= 1-\frac{A_f}{A_r}	\\
h_2 &= 1-\frac{\eta_f}{\eta_r}	\\
h_3 &= 1-\frac{j_f(V^*)}{j_r(V^*)}
\end{align*}  
where $h_1$ measures the ratio of areas under the forward and reverse I-V curves, $h_2$ the ratio of power maxima, and $h_3$ the ratio of currents at a specified voltage. Qualitatively the agreement between these indices and the susceptance is good, but there is clear variance between $h_1$, $h_2$ and $h_3$ showing that the choice of index is not without consequence. A curious feature is that whilst the fullerene passivation clearly reduces hysteresis as measured by EIS, the I-V indices take similar values in both cases. A reduction in the magnitude of hysteresis with passivation is expected, due to its intimate relation with recombination. This may indicate that the susceptance is closer to being a quantity with real physical as well as practical relevance compared to the hysteresis indices. We note that both the susceptance and hysteresis indices are capable of taking on negative values, which is the defining characteristic of ``inverted'' hysteresis discussed extensively in refs. \cite{Tress_Inverted_2016,Shen_Inverted_2017}. Despite their advantages over hysteresis indices, it is important to recognize that susceptance measurements will vary, sometimes quite significantly, with the chosen operating point. This point is discussed further in the supplementary information in relation to Fig. S3, which shows measurements of the variation in susceptance with applied voltage and illumination levels. The results of Fig. S3 demonstrate that the choice of operating point matters: to maximize practical relevance as a measurement of hysteresis, the susceptance should be reported at or near the cell's maximum power point under realistic illumination conditions.

\section*{Conclusion}


Ionic drift-diffusion simulations have been used to interpret the most common features seen in EIS measurements of standard perovskite solar cells. To aid this analysis we have described how the AC capacitance obtained from measurements of a solar cell can be considered as a sum of two components, one from conventional charging currents and the other from phase-delayed recombination current. In perovskite cells we have argued that the contribution of phase-delayed recombination to capacitance measurements, and to EIS more generally, is highly significant. It is therefore mistaken to view all capacitance measurements as an indication of the device's charge-storage capability.  In MAPbI$_3$ ions appear to modulate the recombination of carriers through their influence on the internal field, yielding large additive contributions to the measured capacitance. Other mechanisms can also be envisaged which would require recognizing the dual nature of capacitance. In particular, we presented simulated evidence that a small but noticeable light-induced high-frequency capacitance can result purely from carrier dynamics, with a mechanism that is similar to the ones responsible for producing negative capacitance in conventional diodes \cite{Ershov_Negative_1998}. At low frequencies, we have argued that phase-delayed recombination due to mobile ions is almost certainly responsible for observations of giant photo-induced capacitance. Inductive loop features and measurements of low-frequency negative capacitance can also be explained with these concepts, but will require further studies for experimental confirmation. We have also argued that the standard drift-diffusion models of ion migration must be supplemented in order to properly account for the large ion-electrode capacitance observed experimentally in dark measurements. Here there is also room for further input from experiments on the relevant mechanisms ionic charge compensation at perovskite interfaces.


The success of the models presented here in explaining several major features of interest, such as giant photo-induced capacitance, inductive loop features and capacitance-voltage curves, builds on their earlier achievements in the domain of I-V hysteresis \cite{van_Reenen_Modeling_2015,Richardson_Can_2016,Calado_Evidence_2016,Jacobs_Hysteresis_2017,Shen_Inverted_2017}. The interpretations that follow from these models therefore fit into a unified understanding of transient behaviour, which should enable more integrated usage of EIS as a tool to supplement standard I-V measurements. On this topic we have shown that measurements of susceptance provide a straightforward and relatively unambiguous quantifier of hysteresis, with a simple conceptual interpretation related to the area of the Lissajous ellipse. The adoption of EIS as a method to quantify hysteresis would resolve much of the ambiguity in statements about ``hysteresis-free'' cells, and better distinguish between genuinely suppressed hysteresis as opposed to changes in its characteristic timescale.

\section{Acknowledgement}
This project received funding from the Australian Renewable Energy Agency ARENA and the Australian Centre for Advanced Photovoltaics (ACAP). The views expressed are not necessarily the views of the Australian Government, which does not accept responsibility for any of the information or advice contained herein. 

\section{Supporting Information}
Details on equations, C-V measurements and simulations of the C-V hysteresis, supplementary calculations concerning ionic compensation and geometric capacitance in mesoporous cells.

\bibliography{bib}

\end{document}